\documentclass[10pt,twocolumn]{article}

\usepackage[letterpaper,top=0.9in,bottom=1.0in,left=0.75in,right=0.75in]{geometry}
\usepackage{mathptmx}
\usepackage{amsmath,amssymb}
\usepackage{booktabs}
\usepackage{enumitem}
\usepackage{algorithm}
\usepackage{algpseudocode}
\usepackage{url}
\usepackage[colorlinks=true,allcolors=blue,breaklinks=true]{hyperref}

\floatname{algorithm}{Algorithm}
\setlist{itemsep=1pt,parsep=0pt,topsep=3pt}

\hypersetup{%
	pdfauthor={Ashly Joseph},
	pdftitle={Reducing Hallucinations in Large Language Models through Integrated Self-Verification and Retrieval-Augmented Generation},
	pdfkeywords={Large Language Models, Hallucination Mitigation, Chain-of-Verification, Retrieval-Augmented Generation, Factual Consistency, Trustworthy AI, Explainable AI, Engineering Design Automation, Computer-Aided Design, Standards Compliance},
	pdfsubject={CoVe-RAG+: A Scalable Solution for Trustworthy AI Systems in Engineering},
}

\title{\vspace{-1.2em}\bfseries Reducing Hallucinations in Large Language Models
Through Integrated Self-Verification and Retrieval-Augmented Generation}

\author{%
Ashly Joseph\\[2pt]
\normalsize Cisco Systems, San Jose, CA, USA\\
\normalsize \texttt{ashlyelsy@gmail.com}
}
\date{}

\begin{document}

\makeatletter
\twocolumn[
\begin{@twocolumnfalse}
\maketitle
\begin{abstract}
\noindent
Large Language Models (LLMs) are progressively used for advanced engineering tasks, includes Computer-Aided Design (CAD) documentation, standards compliance verification, and knowledge retrieval. Still, they are prone to produce hallucinations, outputs that seem convincing but aren't based on context that limit their trustworthiness in high-end engineering applications where precision and compliance are crucial. The paper introduces CoVe-RAG+, a unified framework that integrates Chain-of-Verification (CoVe) with Retrieval-Augmented Generation (RAG) to mitigate hallucinations in the results generated by large language models (LLMs). CoVe-RAG+ supports LLM verification in external sources of authority, such as engineering standards, CAD information, and simulation reports, while applying an iterative self-verification process to validate important claims. CoVe-RAG+ is assessed on engineering activities such as CAD model documentation, standards compliance verification, and the reutilization of historical design data. Experimental findings indicate a 28\% improvement in factual accuracy relative to baseline CoVe and RAG methodologies. Moreover, CoVe-RAG+ strengthens user confidence by providing elucidative verification reports and source traceability. The findings indicate that CoVe-RAG+ provides a scalable and reliable option for implementing LLMs in engineering design processes where factual accuracy is critical.
\end{abstract}

\vspace{1.0em}
\end{@twocolumnfalse}
]
\makeatother

\renewcommand{\thefootnote}{}
\footnotetext{Accepted manuscript. Published as: A. Joseph, ``Reducing
Hallucinations in Large Language Models Through Integrated Self-Verification
and Retrieval-Augmented Generation,'' \emph{Proceedings of the ASME 2025
International Design Engineering Technical Conferences and Computers and
Information in Engineering Conference (IDETC/CIE2025)}, Anaheim, CA, USA,
August 17--20, 2025, Paper No. DETC2025-169730, V02BT02A032, 7 pages.
\textcopyright{} 2025 by ASME. The version of record is available at
\url{https://doi.org/10.1115/DETC2025-169730}.}
\renewcommand{\thefootnote}{\arabic{footnote}}

\section{INTRODUCTION}

Large Language Models (LLMs) have shown outstanding abilities in Natural Language Processing (NLP), achieving state-of-the-art results across several tasks like machine translation, question answering, and summarization~\cite{wei2022chain, radford2019language}.
Through extensive pretraining on large textual datasets, LLMs generalize knowledge and adapt to diverse tasks without the need for task-specific adjustments. Despite these advancements, a core difficulty persists: Large Language Models (LLMs) sometimes cause hallucinations~\cite{maynez2020faithfulness}, producing factually inaccurate or fabricated information that seems credible and coherent.

Hallucination undermines the trustworthiness and dependability of AI systems.~\cite{rawte2023survey} This poses enormous challenges in critical fields such as healthcare, law, and engineering design, where incorrect or misleading data can result in unexpected outcomes, safety issues, and financial damage. In engineering design, a hallucinated material property or compliance claim in AI-assisted processes can lead to expensive mistakes, risking product quality and safety.

Engineering design automation increasingly utilizes LLMs for essential functions, such as composing Computer-Aided Design (CAD) model descriptions, assessing compliance to industry standards (e.g. ASME and ISO), and retrieving historical design information for use in new projects.~\cite{makatura2024large} These applications aim to improve design efficiency, decrease manual labor, and help with more informed decision-making.~\cite{abdollahi2025hardware} Nonetheless, the usefulness of LLMs in engineering is hindered by their vulnerability to hallucination, which may lead to non-compliant designs, safety violations, defects in production, and expensive redesign processes.~\cite{rawte2023survey}

Recent advancements have aimed to reduce hallucination in LLMs using two separate ways. Chain-of-Verification (CoVe)~\cite{dhuliawala2023chain} establishes a systematic self-verification mechanism wherein a large language model (LLM) assesses its outputs by asking and reacting to verification inquiries and verifying its own answers. CoVe has shown enhancements in answer dependability; however, it is still limited by the model's internal knowledge and pretraining constraints. Retrieval-Augmented Generation (RAG)~\cite{lewis2020retrieval} provides an alternate method by establishing LLM replies in outside sourced knowledge from trustworthy sources; however, it lacks means for verifying the accurate use of this information.

This study introduces CoVe-RAG+, an effective framework that combines the advantages of CoVe and RAG to reduce hallucinations in AI-assisted engineering processes. CoVe-RAG+ uses external document retrieval to provide important, current knowledge and implements a systematic verification method to ensure that the model's outputs align with the retrieved evidence. By closely integrating retrieval and verification, CoVeRAG+ reduces knowledge gaps and reasoning inaccuracies, producing highly trustworthy AI-assisted solutions for engineering design tasks.
This study demonstrates the effectiveness of CoVe-RAG+ through extensive trials on tasks including CAD model documentation, design compliance validation, and design knowledge retrieval. This research indicates significant improvements in factual correctness, decreased hallucination rates, and increased trust in AI outputs, making CoVe-RAG+ a compelling choice for trustworthy AI-driven engineering design automation.

\section{RELATED WORK}

\subsection{Hallucination in LLMs}

Hallucination is a well-documented problem in large language models, including prominent models such as GPT and BERT~\cite{maynez2020faithfulness}. Although their remarkable fluency, these models are capable of producing factually inaccurate, incomplete, or falsified content, frequently with considerable confidence. Hallucinations generally occur because of insufficient grounding in external sources of knowledge and the constraints of pretraining on static datasets.~\cite{tonmoy2024comprehensive} These challenges are intensified in specialized fields such as engineering, where current and precise subject knowledge is crucial.

Various methods have been suggested for reducing hallucinations.~\cite{huang2023survey} Reinforcement Learning from Human Feedback (RLHF)~\cite{ouyang2022training} fine-tunes large language models (LLMs) by utilizing human preferences to better align outputs with human assessment. Although RLHF enhances response quality, it is insufficient for eliminating hallucinations, particularly in knowledge-intensive activities. Chain-of-Thought (CoT) prompting~\cite{wei2022chain} promotes step-by-step reasoning to improve output consistency and accuracy; nevertheless, its effectiveness declines in domains where large language models (LLMs) possess insufficient knowledge.

\subsection{Chain-of-Verification (CoVe)}

Chain-of-Verification (CoVe)~\cite{dhuliawala2023chain} extends the Chain-of-Thought prompting technique by implementing a formal, multi-step verification framework. CoVe operates through the following stages:
\begin{enumerate}
    \item \textbf{Baseline Response Generation}: The LLM produces an initial draft of the response to a user inquiry.
    \item \textbf{Verification Planning}: The model recognizes essential facts inside its draft and develops verification questions.
    \item \textbf{Verification Execution}: Each verification question is answered independently, utilizing the model's internal knowledge.
    \item \textbf{Final Response Generation}: The preliminary response changes according to verification answers, rectifying any detected errors.
\end{enumerate}

CoVe lowers hallucination rates and enhances factual consistency by prompting the LLM to critically examine its own outputs. CoVe's reliance on the model's internal knowledge is a notable restriction in specialized domains like engineering, where detailed and often updated information is essential. In the absence of external data, CoVe may unknowingly validate incorrect responses.

\subsection{Retrieval-Augmented Generation (RAG)}

Retrieval-Augmented Generation (RAG)~\cite{lewis2020retrieval} reduces the knowledge limitations of large language models (LLMs) by including external document retrieval into the generation process. RAG systems initially collect relevant information from external knowledge repositories, such as private design databases, PDFs or engineering standards, and subsequently inform the LLM's answer based on this collected context. This grounding approach provides LLMs with access to precise, current information outside their training data.

RAG has shown successful in mitigating hallucinations by ensuring that LLM replies are based on trustworthy external sources. Nevertheless, RAG lacks a verification mechanism to ensure that collected evidence is accurately interpreted or utilized. Consequently, LLMs using RAG may still create inaccuracies by altering obtained data or improperly integrating it with their internal knowledge.~\cite{yasunaga2022deep}

\subsection{Integrating CoVe and RAG}

The drawbacks of both CoVe and RAG require the formation of a unified framework that combines their combined advantages. CoVe offers a methodical self-verification technique, whereas RAG anchors replies in external information. The CoVe-RAG+ framework integrates external knowledge retrieval with structured verification, allowing LLMs to provide factually correct and reliable outputs in complex engineering design processes. This method mitigates knowledge gaps and reasoning inaccuracies, providing dependable AI support in activities such as CAD documentation, standards compliance verification, and design knowledge reuse.

\section{METHODOLOGY}

This section introduces the CoVe-RAG+ framework, an adaptive and iterative methodology that combines Chain-of-Verification (CoVe) with Retrieval-Augmented Generation (RAG). CoVe-RAG+ is designed to reduce hallucinations in Large Language Models (LLMs), providing reliable and verifiable outputs suitable for critical engineering design processes, including standards compliance validation and CAD documentation development.

In contrast to current methodologies, CoVe-RAG+ enhances retrieval and verification through three major innovations:
\begin{itemize}
    \item \textbf{Dynamic Re-retrieval}: Triggered by low verification confidence to guarantee comprehensiveness.
    \item \textbf{Multi-modal Data Integration}: Supports textual data, CAD data, and simulation design reports.
    \item \textbf{Explainability Layer}: Delivers transparent verification reports and source traceability to enhance user confidence.
\end{itemize}

\subsection{Overview of CoVe-RAG+}

CoVe-RAG+ integrates two complementary approaches:
\begin{enumerate}
    \item \textbf{Adaptive Retrieval-Augmented Generation (RAG)}: Accurately obtains relevant and context based documents, CAD data, and engineering standards from external knowledge repositories. The retrieval process includes adaptive ranking to prioritize documents based on source authority and semantic relevance.
    \item \textbf{Iterative Chain-of-Verification (CoVe)}: Conducts multi-stage self-verification of the LLM-generated replies, evaluating the truthfulness of the information. The system calculates confidence scores for each validated claim and conducts further retrieval if inconsistencies are detected.
\end{enumerate}

CoVe-RAG+ combines adaptive retrieval with iterative verification to address knowledge incompleteness and reasoning inconsistencies, delivering factually verified replies appropriate for compliance-driven engineering workflows.

\subsubsection{Step 1: Task Classification and Query Processing}

Upon receiving a user query $Q$, the system initially categorizes the kind of question (e.g., compliance verification, CAD documentation, or knowledge retrieval). This categorization informs the ensuing retrieval procedure and validation criteria.

\subsubsection{Step 2: Adaptive Retrieval}

The system retrieves a set of relevant documents $D = \{d_1, d_2, ..., d_k\}$ from external knowledge bases. CoVe-RAG+ supports multi-modal retrieval that encompasses various data types essential for engineering operations. These include textual engineering standards, CAD data and design reports, as well as simulation results when applicable.

Retrieval is executed by a hybrid methodology that integrates dense and sparse search techniques. Dense retrieval uses semantic similarity search techniques, such as FAISS~\cite{johnson2017billion}, to locate documents using on vector embeddings that capture contextual significance. Sparse retrieval, on the other hand, uses keyword matching algorithms like BM25~\cite{robertson1995okapi} to ensure high recall of domain-specific phrases and entities that may be neglected by dense embeddings.

An adaptive scoring system is being used to rate the retrieved documents efficiently. Two primary factors govern this ranking procedure. Initially, the legitimacy of sources is assessed, giving priority to documents from reputable and authoritative entities rather than community-sourced or poorly reviewed resources. Secondly, newness is considered, prioritizing more recently released documents in fields where knowledge progresses quickly. This adaptive retrieval procedure ensures that the document set $D$ is both precise and contextually pertinent, establishing a dependable basis for subsequent answer generation and verification stages.

\subsubsection{Step 3: Baseline Response Generation}

The LLM produces an initial response $A_0$ based in the obtained documents $D$. The baseline response comprises:
\begin{itemize}
    \item Extractive elements: Direct references to source documents
    \item Abstractive elements: Consolidated insights relevant to the user query
\end{itemize}

\subsubsection{Step 4: Iterative Chain-of-Verification}

The system detects factual claims $C = \{c_1, c_2, ..., c_n\}$ in $A_0$. For each claim $c_i$, the following verification process is conducted:
\begin{enumerate}
    \item Generate verification question $VQ_i$
    \item Answer $VQ_i$ using the document set $D$
    \item Compute a confidence score $S_i$ for the answer
    \item If $S_i$ is below a pre-defined threshold:
    \begin{enumerate}
        \item Retrieve additional documents $D'$
        \item Update $D \gets D \cup D'$
        \item Re-answer $VQ_i$ using the updated $D$
    \end{enumerate}
\end{enumerate}

Explainable verification traces are produced for each claim, which include:
\begin{itemize}
    \item The justification for the verification result
    \item Source documents utilized as evidence
\end{itemize}

\subsubsection{Step 5: Final Verified Response Synthesis}

The final confirmed response $A_f$ is assembled by combining $A_0$ with validated claims $VA = \{VA_1, VA_2, ..., VA_n\}$. Unverifiable claims are flagged for user evaluation. A verification report is produced, outlining:
\begin{itemize}
    \item Confidence scores for each validated claim
    \item Verified and flagged claims with supporting proof
\end{itemize}

\subsection{Algorithmic Workflow}

The algorithmic workflow of CoVe-RAG+ is described in Algorithm~\ref{alg:cove-rag}.

\begin{algorithm}[htbp]
\caption{CoVe-RAG+ Workflow}
\label{alg:cove-rag}
\begin{algorithmic}[1]
\Require User Query $Q$
\Ensure Verified Response $A_f$ and Explainability Report $R$

\State Classify $Q$ to determine the task type $T$
\State Retrieve documents $D \gets \text{Retrieve}(T, Q)$
\State Produce baseline response $A_0 \gets \text{LLM}(D, Q)$
\For{each claim $c_i$ in $A_0$}
    \State Generate verification question $VQ_i$
    \State Answer $VQ_i$ to obtain $VA_i$ using $D$
    \State Calculate confidence score $S_i$
    \If{$S_i < \text{Threshold}$}
        \State Retrieve additional documents $D'$
        \State Update $D \gets D \cup D'$
        \State Re-evaluate $VQ_i$ using updated $D$
    \EndIf
\EndFor
\State Compile $A_f$ from verified claims $VA$
\State Produce explainability report $R$
\State \Return $A_f$, $R$
\end{algorithmic}
\end{algorithm}

\subsection{Architecture Components}

The CoVe-RAG+ architecture has six fundamental components:
\begin{itemize}
    \item \textbf{Task Classifier}: Identifies the query type to guide retrieval and verification.
    \item \textbf{Multi-modal Document Retriever}: Extracts textual data, CAD metadata, and simulation reports with hybrid retrieval techniques.
    \item \textbf{LLM Generator}: Generates baseline responses, verification questions, and verification answers. Fine-tuned and optimized for engineering contexts.
    \item \textbf{Verifier Module}: Conducts iterative verification with confidence scoring and dynamic retrieval triggers.
    \item \textbf{Explainability Layer}: Produces user-friendly verification logs and explanations.
    \item \textbf{Response Synthesizer}: Generates coherent, compliance-ready answers by integrating verified context content.
\end{itemize}

\subsection{CoVe-RAG+: Key Innovations and Differentiators}

CoVe-RAG+ differentiates itself from prior CoVe-RAG frameworks by this three key innovations:
\begin{itemize}
    \item \textbf{Multi-modal Adaptive Retrieval:} Unlike prior methods that focus on text-based retrieval, CoVe-RAG+ retrieves a variety of data kinds, such as CAD component information and simulation reports, hence improving contextual relevance in engineering fields.
    \item \textbf{Dynamic Re-retrieval Trigger:} Verification confidence scores facilitate dynamic re-retrieval, ensuring that claims having low-confidence prompt further document retrieval and re-verification.
    \item \textbf{Explainability Layer:} CoVe-RAG+ features a verification report generator that offers end-users clear and verifiable proof for every factual claim provided in the final response.
\end{itemize}

\section{EXPERIMENTAL SETUP}

This section analyzes the efficiency of the CoVe-RAG+ framework in engineering design tasks, emphasizing actual correctness, hallucination prevention, and compliance to domain-specific standards.

\subsection{Datasets}

Experiments were performed on two multi-modal datasets relevant to engineering and industrial applications. The initial dataset, Wikidata Engineering Subset~\cite{wikidata2021}, is a curated collection that focuses on manufacturing entities, engineering standards, and essential engineering concepts. A carefully chosen set of data from Wikidata that focuses on industrial companies, engineering standards, and mechanical parts. Entities were sorted by groups that were useful, like ``Mechanical Parts'' and ``Material Properties,'' and subject experts checked that the results were correct. There are 45,000 individuals and 80,000 triples in the collection. Because of licensing restrictions, the dataset can only be used for research reasons if asked for.
The second dataset, MultiSpanQA (Engineering Edition)~\cite{xiong2023multispanqa}, is made up of factoid-based question-answer pairs that have been enhanced with engineering knowledge and multi-modal references like CAD design specs and modeling reports.

The GrabCAD CAD Metadata Collection~\cite{grabcad2024} has 50,000 open-source CAD files with 3D models (STEP, IGES), component specs, design notes, and material attributes. These files can be used with the GrabCAD license. The ArXiv Engineering Papers Corpus~\cite{arxiveng2024} is made up of 15,000 open-access engineering papers from arXiv that have had their abstracts, sections, and figures taken out that are useful for jobs in robots, aircraft, materials, and industrial systems. The NIST Materials Data Repository (MDR)~\cite{nistmdr2024} has 30,000 directly measured and generated material datasets for metals, polymers, and composites. These datasets can be accessed through the NIST MDR application.

\subsection{Models and Tools}

The CoVe-RAG+ architecture was executed with advanced language models and retrieval tools specifically designed for engineering applications. Mistral 7B~\cite{mistral2023open} and LLaMA 2 65B~\cite{touvron2023llama} models were fine-tuned on engineering-specific datasets to improve domain knowledge and terminology correctness.

LangChain~\cite{langchain2023docs} has been integrated with ElasticSearch to enable both dense and sparse retrieval across structured and unstructured data sources. Vector indexing was handled by FAISS~\cite{johnson2017billion} to provide high-performance similarity searches, while BM25~\cite{robertson1995okapi} was used to guarantee efficient keyword retrieval. The system also included modules for parsing CAD information and integrating simulation reports, offering extensive multi-modal support within the retrieval and verification process.

\subsection{Training Details}
CoVe-RAG+ was run in PyTorch. Mistral-7B was fine-tuned using Low-Rank Adaptation (LoRA)~\cite{hu2021lora} on 40,000 domain-specific question-answer pairs. Training was conducted with the batch size of 512, over 5 epochs, with an AdamW optimizer~\cite{loshchilov2019decoupled} and a learning rate of $5 \times 10^{-5}$. The dynamic re-retrieval module was developed using LangChain and FAISS for adaptive document retrieval and ranking.

\subsection{Evaluation Metrics}

The examination of CoVe-RAG+ was conducted using a combination of automated evaluation and human evaluations. Precision and recall were assessed to evaluate the accuracy of fact retrieval and the system's capacity to reduce hallucinations in produced responses. FACTSCORE~\cite{min2023factscore} was utilized to evaluate sentence-level factual consistency, providing an objective measure of accuracy. In addition, human evaluation was conducted by domain experts by checking generated outputs for factual truth, readability, and compliance with engineering terms. This made sure that the system could be used for important engineering tasks.

\subsection{Experimental Procedure}

Each model had to come up with answers to a set of predefined technical questions. The verification process made sure that facts were checked over and over again, and dynamic re-retrieval was used when confidence levels were not met.

All tests were conducted on a server with:
\begin{itemize}
    \item Dual NVIDIA A100 GPUs
    \item 512GB RAM
    \item 64-core AMD EPYC processor
\end{itemize}

\section{RESULTS AND DISCUSSION}

\subsection{Quantitative Results}

Table~\ref{tab:results} explores the performance comparison of the baseline CoVe, conventional RAG, and the proposed CoVe-RAG+ framework.

\begin{table}[htbp]
\caption{Performance Comparison on Engineering Datasets}
\label{tab:results}
\centering
\begin{tabular}{@{}lccc@{}}
\toprule
\textbf{Model} & \textbf{Precision} & \textbf{Recall} & \textbf{FACTSCORE} \\ \midrule
Baseline CoVe  & 0.36      & 0.38   & 55.9 \\
RAG Only       & 0.40      & 0.46   & 60.8 \\
CoVe-RAG+     & \textbf{0.48}      & \textbf{0.50}   & \textbf{71.4} \\ \bottomrule
\end{tabular}
\end{table}

\subsection{Human Evaluation}
A human review with five mechanical engineering experts was done to check the true correctness and usefulness of the CoVe-RAG+ system outputs. Each expert had more than ten years of experience in things like checking designs, making sure they follow the rules, and documenting designs in CAD. The judges looked at a sample of 100 answers on their own, focused on three main factors: (1) correctness of the facts, (2) compliance with technical standards, and (3) the system's explanations being clear and logical.

Cohen's Kappa was used to measure inter-rater dependability, and a score of 0.82 meant that there was strong agreement~\cite{cohen1960coefficient}. Table~\ref{tab:human} reports the average scores across evaluators.

\begin{table}[htbp]
\caption{Human Evaluation Scores (Average Across Evaluators)}
\label{tab:human}
\centering
\begin{tabular}{@{}lcc@{}}
\toprule
\textbf{Metric} & \textbf{Baseline CoVe} & \textbf{CoVe-RAG+} \\ \midrule
Factual Accuracy & 3.8 / 5 & 4.6 / 5 \\
Compliance       & 3.5 / 5 & 4.4 / 5 \\
Explainability   & 3.0 / 5 & 4.5 / 5 \\ \bottomrule
\end{tabular}
\end{table}

\subsection{Statistical Significance Analysis}

Statistical significance tests were used to verify improvements in factual correctness and compliance.

\begin{itemize}
    \item \textbf{T-Test Results:} The enhancements in factual accuracy between CoVe-RAG+ and Baseline CoVe are statistically significant ($p$-value $<$ 0.01).
    \item \textbf{Confidence Intervals:} CoVe-RAG+ achieved a 95\% confidence interval ranging from 4.4 to 4.8 for factual correctness.
\end{itemize}

\subsection{Error Analysis}

An error analysis was conducted on wrong outputs from CoVe-RAG+. Main findings include:
\begin{itemize}
    \item \textbf{Ambiguous Queries:} 12\% of hallucinations occurred in queries lacking clearly defined specifications.
    \item \textbf{Outdated Standards:} 8\% mistakes originated from the utilization of outdated standards documents during retrieval.
    \item \textbf{CAD Metadata Misinterpretation:} 6\% involved incorrect understanding of CAD information.
\end{itemize}

These findings indicate opportunities for future improvement, including the integration of real-time standard updates and the optimization of CAD metadata parsing methods.

\subsection{System Performance and Scalability}

To ensure industrial scalability, CoVe-RAG+ incorporates:
\begin{itemize}
    \item \textbf{Parallel Retrieval and Verification Pipelines}: Minimizes overall delay by simultaneously executing verification operations.
    \item \textbf{Optimized Batch Sizes}: Adaptive batching for retrieval and verification balances computational expense and response time.
    \item \textbf{Threshold Tuning}: The dynamic modification of confidence thresholds reduces unnecessary re-retrievals.
\end{itemize}

\section{CONCLUSION}

The paper introduces CoVe-RAG+, an adaptive and iterative approach that combines Chain-of-Verification (CoVe) with Retrieval-Augmented Generation (RAG) to reduce hallucinations in Large Language Models (LLMs). CoVe-RAG+ utilizes dynamic re-retrieval, multi-modal data integration, and explainability mechanisms to produce trustworthy and verifiable outputs, effectively tackling significant difficulties in high-stakes engineering design workflows, including standards compliance and CAD documentation.
Experimental assessments indicate that CoVe-RAG+ enhances factual correctness and consistency relative to baseline CoVe and RAG methodologies, achieving a 28\% improvement in precision as well as improved FACTSCORE measures. The system offers scalable, transparent AI solutions that may be used to various fields requiring strong factual integrity.

\section{LIMITATIONS}

There are some things that the CoVe-RAG+ system can't do, even though it makes Large Language Models (LLMs) much more truly accurate. The multiple-stage verification and dynamic retrieval methods add extra computation to the machine, which makes it harder to scale and causes delays, especially in real-time applications. The system relies on outside knowledge sources, which can be problematic because data that is out of date, missing, or inconsistent can make verification less accurate.
More than that, CoVe-RAG+ has only been fully tried in engineering domains. Its usability and effectiveness in other high-stakes industries, like healthcare or law, still need to be completely tested.

\section{FUTURE WORK}

Future research will focus on enhancing retrieval and verification procedures via new indexing approaches, parallelization, and parameter-efficient fine-tuning methods like Low-Rank Adaptation (LoRA). Enhancing multi-modal capabilities to manage intricate data types, such as 3D CAD models, simulation data, and visual media, would extend the framework's application. Furthermore, domain adaption and generalization in sectors like as healthcare, aerospace, and legal compliance will be advanced by using domain-specific knowledge graphs and ontologies. Improvements to the explainability layer are expected, encompassing interactive visuals and simplified reports to enhance user understanding and trust.

\end{document}